# Application of Eckart-Hellmann potential to study selected diatomic molecules using Nikiforov-Uvarov-Functional Analysis (NUFA) method

E. P. Inyang[1,2], E. S. William[2], E. Omugbe[3], E. P. Inyang[2], E.A.Ibanga[1], F.Ayedun[1], I.O.Akpan[2] and J.E.Ntibi[2]

[1]*Department of Physics, National Open University of Nigeria, Jabi, Abuja*

[2]*Theoretical Physics Group, Department of Physics, University of Calabar, P.M.B 1115, Calabar, Nigeria*

[3]*Department of Physics, Federal University of Petroleum Resources,Effurun, Delta State,Nigeria*

*Corresponding author email: etidophysics@gmail.com OR einyang@noun.edu.ng*

**Abstract**

The energy levels of the Schrödinger equation under the Eckart-Hellmann potential (EHP) energy function are studied by the Nikiforov-Uvarov-Functional Analysis (NUFA) method. We obtained the analytic solution of the energy spectra and the wave function in closed form with the help of Greene-Aldrich approximation. The numerical bound states energy for various screening parameters at different quantum states and vibrational energies of EHP for CuLi, TiH, VH, and TiC diatomic molecules were computed. Four particular cases of this potential were achieved. To test the accuracy of our results, we computed the bound states energy eigenvalues of Hellmann potential which are in excellent agreement with the report of other researchers.

1. Introduction

The analytical methods for solving bound state problems that arise in physics and their applications have received much attention over the years. The development of these methods allows one to derive the analytic eigen-solutions of the relativistic and non-relativistic wave equations which play a crucial role in interpreting the behavior of quantum mechanical systems. The frequently used analytical methods are the Nikiforov-Uvarov method (NU) [1-30], Asymptotic iterative method (AIM) [31], Laplace transformation approach [32], ansatz solution method [33], super-symmetric quantum mechanics approach (SUSYQM) [34,35], exact and proper quantization methods [36,37], series expansion method [38-45], the recent study via the Heun function approach has been used widely to study those soluble quantum systems which could not be solved before,e.g. the systems including the Mathieu potential,rigid rotor problem,sextic type problem, Konwent potential and others[46-54]

The Schrödinger equation (SE) can be studied for different quantum–mechanical processes with the above analytical methods [55-58]. The analytical solutions to this equation with a physical potential plays an important role in our understanding of the fundamental root of a quantum system. This is because the eigenvalues and eigenfunctions contain vital information concerning the quantum system under study [59, 60]. However, the exact bound state solutions of the SE of a number of these potentials are possible in some cases for example, Coulomb potential [61]. Obtaining the approximate solutions when the arbitrary angular momentum quantum number $l$ is not equal to zero, one can solve the SE utilizing a reasonable approximation schemes, like the Pekeris, Greene and Aldrich, and others [62-67].

The Eckart potential [68], presented by Eckart in 1930, is a diatomic molecular potential model. Because of its significance in physics and chemical physics, numerous authors in references therein [69-74], considered the bound state solutions of the wave equations for this potential.

Hellmann potential [75], has been widely utilized by numerous authors to obtain bound state solutions in atomic, nuclear, and particle physics [76-79]. The Hellmann potential finds application in condensed matter physics [80].

As of late, many researchers have shown a great deal of interest towards the combination of at least two potentials. The essence of joining at least two physical potential models is to take into account more physical application and comparative analysis to existing investigations of molecular physics [81-85].

With this in mind, we aim at obtaining the approximate bound state analytical solutions to the SE with the Eckart plus Hellmann potential using Nikiforov-Uvarov-Functional Analysis (NUFA) method. The obtained energy equation will be applied to study the energy spectra of some selected diatomic molecules. The combined potential takes the form [68,75]

$$V(r) = -\frac{Ae^{-\alpha r}}{1-e^{-\alpha r}} + \frac{Be^{-\alpha r}}{\left(1-e^{-\alpha r}\right)^2} - \frac{C}{r} + \frac{De^{-\alpha r}}{r}, \qquad (1)$$

where A, B, C and D are the strength of the potential, $\alpha$ is the screening parameter and $r$ is inter molecular distance.

The paper is organized as follows: In Sect. 2, a brief introduction of the NUFA method is presented. In Sect. 3 we solve the SE with the EHP to obtain the energy equation and wave function. In Sect. 4, the derived energy equation will be used to obtain the numerical computation of energy eigenvalues at different states and selected diatomic molecules. In Sect. 4, we present the results and discussion. Conclusions are given in Sect. 5.

## 2. NIKIFOROV-UVAROV-FUNCTIONAL ANALYSIS (NUFA) METHOD

Using the concepts of the NU, parametric NU and the functional analysis methods [1,86, 87], Ikot et al [88] proposed a simple and elegant method for solving a second order differential equation of the hypergeometric type called Nikiforov-Uvarov-Functional Analysis method (NUFA) method. This method is easy and simple. The NU method is used to solve a second-order differential equation of the form [1]

$$\psi''(z) + \frac{\tilde{\tau}(z)}{\sigma(z)}\psi'(z) + \frac{\tilde{\sigma}(z)}{\sigma^2(z)}\psi(z) = 0 \qquad (2)$$

where $\tilde{\sigma}(z)$ and $\sigma(z)$ are polynomials of maximum second degree and $\tilde{\tau}(z)$ is a polynomial of maximum first degree. Tezcan and Sever [86] latter introduced the parametric form of NU method in the form

$$\psi'' + \frac{\alpha_1 - \alpha_2 z}{z(1-\alpha_3 z)}\psi' + \frac{1}{z^2(1-\alpha_3 z)^2}\left[-\xi_1 z^2 + \xi_2 z - \xi_3\right]\psi(s) = 0 \qquad (3)$$

where $\alpha_i$ and $\xi_i (i=1,2,3)$ are all parameters. It can be observed in Eq. (3) that the differential equation has two singularities at $z \to 0$ and $z \to 1$, thus we take the wave function in the form,

$$\psi(z) = z^{\lambda} (1-z)^{\nu} f(z) \tag{4}$$

Substituting Eq.(4) into Eq.(3) leads to the following equation,

$$z(1-\alpha z)f''(z) + \left[\alpha_1 + 2\lambda - (2\lambda\alpha_3 + 2\nu\alpha_3 + \alpha_2)z\right]f'(z)$$

$$-\alpha_3 \left( \lambda + \nu + \frac{\alpha_2}{\alpha_3} - 1 + \sqrt{\left(\frac{\alpha_2}{\alpha_3} - 1\right)^2 + \frac{\xi_1}{\alpha_3}} \right)$$

$$\times \left( \lambda + \nu + \frac{\alpha_2}{\alpha^2_3} - 1 + \sqrt{\left(\frac{\alpha_2}{\alpha_3} - 1\right)^2 + \frac{\xi_1}{\alpha^2_3}} \right) + \left[ \frac{\lambda(\lambda-1) + \alpha_1\lambda - \xi_3}{z} + \frac{\alpha_2\nu - \alpha_1\alpha_3\nu + \nu(\nu-1)\alpha_3 - \frac{\xi_1}{\alpha_3} + \xi_2 - \xi_3\alpha_3}{(1-\alpha_3 z)} \right] f(z) = 0 \tag{5}$$

Equation (5) can be reduced to a Gauss hypergeometric equation if and only if the following functions are gone,

$$\lambda(\lambda-1) + \alpha_1\lambda - \xi_3 = 0 \tag{6}$$

$$\alpha_2\nu - \alpha_1\alpha_3\nu + \nu(\nu-1)\alpha_3 - \frac{\xi_1}{\alpha_3} + \xi_2 - \xi_3\alpha_3 = 0 \tag{7}$$

Thus Eq.(5) becomes

$$z(1-\alpha_1 z)f''(z) + \left[\alpha_1 + 2\lambda - (2\lambda\alpha_3 + 2\nu\alpha_3 + \alpha_2)z\right]f'(z)$$

$$-\alpha_3 \left( \lambda + \nu + \frac{\alpha_2}{\alpha_3} - 1 + \sqrt{\left(\frac{\alpha_2}{\alpha_3} - 1\right)^2 + \frac{\xi_1}{\alpha_3}} \right) \tag{8}$$

$$\times \left( \lambda + \nu + \frac{\alpha_2}{\alpha^2_3} - 1 + \sqrt{\left(\frac{\alpha_2}{\alpha_3} - 1\right)^2 + \frac{\xi_1}{\alpha^2_3}} \right) f(z) = 0$$

Solving Eqs. (6) and (7) gives Eqs. (9) and (10),

$$\lambda = \frac{(1-\alpha_1)}{2} \pm \frac{1}{2}\sqrt{(1-\alpha_1)^2 + 4\xi_3} \tag{9}$$

$$\nu = \frac{(\alpha_3 + \alpha_1\alpha_3 - \alpha_2) \pm \sqrt{(\alpha_3 + \alpha_1\alpha_3 - \alpha_2)^2 + \left(\frac{\xi_1}{\alpha_3} + \alpha_3\xi_3 - \xi_2\right)}}{2} \tag{10}$$

Equation (8) is the hyper geometric equation type of the form,

$$x(1-x)f''(x)+\left[c+(a+b+1)x\right]f'(x)-abf(x)=0 \tag{11}$$

Using Eqs.(4),(8) and (11), we obtain the energy equation and the corresponding wave equation respectively for the NUFA method as follows:

$$\lambda^2+2\lambda\left(v+\frac{\alpha_2}{\alpha_3}-1+\frac{n}{\sqrt{\alpha_3}}\right)+\left(v+\frac{\alpha_2}{\alpha_3}-1+\frac{n}{\sqrt{\alpha_3}}\right)^2-\left(\frac{\alpha_2}{\alpha_3}-1\right)^2-\frac{\xi_1}{\alpha_3^2}=0 \tag{12}$$

$$\psi(z)=N_z\frac{(1-\alpha_1)+\sqrt{(1-\alpha_1)^2+4\xi_3}}{2}$$

$$\times(1-\alpha_3 z)\frac{(\alpha_3+\alpha_1\alpha_3-\alpha_2)\pm\sqrt{(\alpha_3+\alpha_1\alpha_3-\alpha_2)^2+\left(\frac{\xi_1}{\alpha_3^2}+\alpha_3\xi_3-\xi_2\right)}}{2}\times {}_2F_1(a,b,c;z) \tag{13}$$

where $a, b,$ and $c$ are given as follows;

$$a=\sqrt{\alpha_3}\left(\lambda+v+\frac{\alpha_2}{\alpha_3}-1+\sqrt{\left(\frac{\alpha_2}{\alpha_3}-1\right)^2+\frac{\xi_1}{\alpha_3}}\right) \tag{14}$$

$$b=\sqrt{\alpha_3}\left(\lambda+v+\frac{\alpha_2}{\alpha_3}-1-\sqrt{\left(\frac{\alpha_2}{\alpha_3}-1\right)^2+\frac{\xi_1}{\alpha_3}}\right) \tag{15}$$

$$c=\alpha_1+2\lambda \tag{16}$$

### 3. Approximate solutions of the Schrödinger equation with Eckart plus Hellmann potential

The SE takes the form [2]

$$\frac{d^2 U(r)}{dr^2}+\left[\frac{2\mu}{\hbar^2}(E_{nl}-V(r))-\frac{l(l+1)}{r^2}\right]U(r)=0 \tag{17}$$

$E_{nl}$ is the energy eigenvalues of the quantum system, $l$ is the angular momentum quantum number, $\mu$ is the reduced mass of the system, $\hbar$ is the reduced Planck's constant and $r$ is radial distance from the origin.

Equation (17) cannot be solved exactly with the proposed potential. So we introduce an approximation scheme proposed by Greene-Aldrich [62] to deal with the centrifugal barrier. This approximation is a good approximation to the centrifugal term which is valid for $\alpha\ll 1,$ and it becomes

$$\frac{1}{r^2}\approx\frac{\alpha^2}{\left(1-e^{-\alpha r}\right)^2}. \tag{18}$$

Substituting Eqs. (1) and (18) into Eq.(17), Eq.(19) is obtain as;

$$\frac{d^2 U(r)}{dr^2} + \left[ \frac{2\mu}{\hbar^2} \left( E_{nl} - \frac{Ae^{-\alpha r}}{1-e^{-\alpha r}} - \frac{Be^{-\alpha r}}{\left(1-e^{-\alpha r}\right)^2} + \frac{C\alpha}{1-e^{-\alpha r}} - \frac{De^{-\alpha r}}{1-e^{-\alpha r}} \right) + \frac{l(l+1)\alpha^2}{\left(1-e^{-\alpha r}\right)^2} \right] U(r) = 0 \qquad (19)$$

By using the change of variable from $r \to x$, our new coordinate becomes

$$x = e^{-\alpha r}. \qquad (20)$$

We substitute Eq. (20) into Eq. (19) and after some simplifications; Eq. (21) is gotten as,

$$\frac{d^2 U(x)}{dx^2} + \frac{1-x}{x(1-x)} \frac{dU(x)}{dx} + \frac{1}{x^2(1-x)^2} \left[ -(\varepsilon - \beta_0 - \beta_3) x^2 + (2\varepsilon - \beta_0 + \beta_1 - \beta_2 - \beta_3) x - (\varepsilon - \beta_2 + \gamma) \right] U(x) = 0, \qquad (21)$$

where

$$-\varepsilon = \frac{2\mu E_{nl}}{\alpha^2 \hbar^2}, \quad \beta_0 = \frac{2\mu A}{\alpha^2 \hbar^2}, \quad \beta_1 = \frac{2\mu B}{\alpha^2 \hbar^2}, \quad \beta_2 = \frac{2\mu C}{\alpha \hbar^2}, \quad \beta_3 = \frac{2\mu D}{\alpha \hbar^2}, \quad \gamma = l(l+1) \Big\}. \qquad (22)$$

Comparing Eq. (21) and Eq. (3), we obtain the relevant polynomials as:

$$\alpha_1 = \alpha_2 = \alpha_3 = 1, \xi_1 = \varepsilon - \beta_0 - \beta_3, \xi_2 = 2\varepsilon - \beta_0 + \beta_1 - \beta_2 - \beta_3, \xi_3 = \varepsilon - \beta_2 + \gamma \Big\} \qquad (23)$$

Inserting the polynomials given by Eq. (23) into Eqs. (9) and (10), we have

$$\lambda = \sqrt{4(\varepsilon - \beta_2 + \gamma)} \qquad (24)$$

$$v = \frac{1}{2} \pm \sqrt{1 + 4(\gamma - \beta_1)} \qquad (25)$$

Substituting Eqs. (23), (24),(25) and (22) into Eq.(12), we obtain the energy equation of the EHP as;

$$E_{nl} = \frac{\alpha^2 \hbar^2 (l+l^2)}{2\mu} - A - C\alpha - \frac{\hbar^2 \alpha^2}{8\mu} \left[ \frac{\left( n + \frac{1}{2} + \sqrt{\frac{1}{4} + \frac{2B\mu}{\alpha^2 \hbar^2} + (l+l^2)} \right)^2 - \frac{2A\mu}{\alpha^2 \hbar^2} + \frac{2B\mu}{\alpha^2 \hbar^2} - \frac{2C\mu}{\alpha \hbar^2} + \frac{2D\mu}{\alpha \hbar^2} + (l+l^2)}{n + \frac{1}{2} + \sqrt{\frac{1}{4} + \frac{2B\mu}{\alpha^2 \hbar^2} + (l+l^2)}} \right]^2 \qquad (26)$$

### 3.1 Particular case

To test for the accuracy of our results, we set some parameters in Eqs. (1) and (26) to zero and obtain four particular cases of potential and energy equation.

1. We set $A = B = 0$ and obtain the Hellmann potential and its energy equation as shown in Eqs. (27) and (28)

$$V(r) = -\frac{C}{r} + \frac{De^{-\alpha r}}{r}.$$

(27)

$$E_{nl} = \frac{\alpha^2 \hbar^2 (l+l^2)}{2\mu} - C\alpha - \frac{\hbar^2 \alpha^2}{8\mu} \left[ \frac{\left(n+\frac{1}{2}+\sqrt{\frac{1}{4}+(l+l^2)}\right)^2 - \frac{2C\mu}{\alpha \hbar^2} + \frac{2D\mu}{\alpha \hbar^2} + (l+l^2)}{n+\frac{1}{2}+\sqrt{\frac{1}{4}+(l+l^2)}} \right]^2$$

(28)

Equation (28) is in agreement with Eq. (38) of Ref. [77]

2. We set $C = D = 0$ and obtain the Eckart potential and its energy equation in the form of Eqs. (29) and (30)

$$V(r) = -\frac{Ae^{-\alpha r}}{1-e^{-\alpha r}} + \frac{Be^{-\alpha r}}{(1-e^{-\alpha r})^2}.$$

(29)

$$E_{nl} = \frac{\alpha^2 \hbar^2 (l+l^2)}{2\mu} - A - \frac{\hbar^2 \alpha^2}{8\mu} \left[ \frac{\left(n+\frac{1}{2}+\sqrt{\frac{1}{4}+\frac{2B\mu}{\alpha^2 \hbar^2}+(l+l^2)}\right)^2 - \frac{2A\mu}{\alpha^2 \hbar^2} + \frac{2B\mu}{\alpha^2 \hbar^2} + (l+l^2)}{n+\frac{1}{2}+\sqrt{\frac{1}{4}+\frac{2B\mu}{\alpha^2 \hbar^2}+(l+l^2)}} \right]^2$$

(30)

3. We set $A = B = D = \alpha = 0$, the Coulomb potential and its energy equation in the form of Eqs. (31) and (32) is obtain

$$V(r) = -\frac{C}{r}.$$

(31)

$$E_{nl} = -\frac{\mu C^2}{2\hbar^2 (n\_r+l+1)^2}$$

(32)

Where $n\_r+l+1 = n$ is the principal quantum number.

The result of Eq. (32) is consistent with the result obtained in Eq. (36) in Ref. [21]

4. We set $A = B = C = 0$, the Yukawa potential and its energy equation of the form of Eqs. (33) and (34) is obtain

$$V(r) = \frac{De^{-\alpha r}}{r}.$$

(33)

$$E_{nl} = \frac{\alpha^2 \hbar^2 \left(l+l^2\right)}{2\mu} - \frac{\hbar^2 \alpha^2}{8\mu} \left[ \frac{\left(n+\frac{1}{2}+\sqrt{\frac{1}{4}+\left(l+l^2\right)}\right)^2 + \frac{2D\mu}{\alpha \hbar^2} + \left(l+l^2\right)}{n+\frac{1}{2}+\sqrt{\frac{1}{4}+\left(l+l^2\right)}} \right]^2 \qquad (34)$$

The result of Eq. (34) is consistent with the result obtained in Eq. (38) in Ref. [24]

The corresponding wavefunction is given as

$$\psi(x) = N \frac{\sqrt{4(\varepsilon-\beta_2+\gamma)}}{2} (1-z) \sqrt{4(4(\gamma-\beta_1))} \,_2F_1(a,b,c;z) \qquad (35)$$

where,

$$a = \sqrt{4(\varepsilon-\beta_2+\gamma)} + \frac{1}{2} \pm \sqrt{1+4(\gamma-\beta_1)} + \sqrt{\varepsilon-\beta_0-\beta_3} \qquad (36)$$

$$b = \sqrt{4(\varepsilon-\beta_2+\gamma)} + \frac{1}{2} \pm \sqrt{1+4(\gamma-\beta_1)} - \sqrt{\varepsilon-\beta_0-\beta_3} \qquad (37)$$

$$c = 1 + 2\sqrt{4(\varepsilon-\beta_2+\gamma)} \qquad (38)$$

## 4. Results and discussion

To test the accuracy of our results, we computed the bound states energy eigenvalues of EHP with $\hbar = \mu = 1$ using arbitrary potential parameters as presented in table 1. The result shows, as the screening parameter and potential strength increases, there is a decrease in energy eigenvalues at different quantum states. We apply the experimental data obtained from Ref. [89] as presented in table 2 and also, adopted the conversions: $1 \text{ amu} = 931.494028 \text{ MeV}/c^2$ and $\hbar c = 1973.29 \text{ eV Å}$ [90] to compute the vibrational energies of EHP for CuLi, TiH, VH, and TiC diatomic molecules using Eq. (26). The numerical computation is given in Tables 3. It is observed that for each vibrational quantum number, the vibrational energies increase with increase in the rotational quantum number, for each of the selected diatomic molecules.

The numerical energy eigenvalues for Hellmann potential is also computed to check for the accuracy of the NUFA method as presented in Table 4. The result is in good agreement with the earlier results of Ref. [76] with NU, AP, and SUSY method of [77] and PT method of [78].

From Figs. 1(a) and (b) – Figs. 4(a) and (b) respectively, we plotted the ground and excited states energy eigenvalues of the different quantum states as a function of the EHP strengths, respectively. We observed that there is a decrease in energy in both the ground and excited states as the potential strength, A, B, C and D increases, respectively. In Fig. 5(a) and 5(b), we plotted the energy eigenvalues of EHP versus the screening parameter. Here, the energy increases for $\ell = 0, 1, 2$ and $3$ in the ground states and decreases in $\ell = 4$, as the screening parameter increases. We also observed increase in energy for $\ell = 0, 1,$ and $2$ and decrease in $\ell = 3$ and $4$ as the screening parameter increases in the excited states.

TABLE 1. Bound state energy eigenvalues $(eV)$ of the Eckart-Hellmann potential with $\hbar = \mu = 1$

| State | $\alpha$ | $A = 0.01, B = 0.5,$ $C = 1, D = -1$ | $A = 0.005, B = 0.25,$ $C = 2, D = -2$ | $A = 0.0025, B = 0.125,$ $C = 4, D = -4$ |
|---|---|---|---|---|
| 1s | 0.025 | -0.5263521625 | -0.3104395873 | -0.3041019021 |
|  | 0.050 | -0.5563594522 | -0.3923969578 | -0.6301704138 |
|  | 0.075 | -0.5898519860 | -0.4938113093 | -1.081923010 |
|  | 0.100 | -0.6266198628 | -0.6127403461 | -1.640204160 |
|  | 0.150 | -0.709198628 | -0.8960512898 | -3.009546904 |
| 2s | 0.025 | -0.5278462720 | -0.3134850809 | -0.3064758005 |
|  | 0.050 | -0.5622260539 | -0.4015176484 | -0.6147211738 |
|  | 0.075 | -0.6021935108 | -0.5081767351 | -1.004551117 |
|  | 0.100 | -0.6469257426 | -0.6289588255 | -1.445599734 |
|  | 0.150 | -0.7480659625 | -0.8999476994 | -2.405459656 |
| 2p | 0.025 | -0.5257258961 | -0.3097912239 | -0.3031052053 |
| | 0.050 | -0.5538354335 | -0.3895266085 | -0.6221624598 |
| | 0.075 | -0.5841024422 | -0.4863981654 | -1.050909737 |
| | 0.100 | -0.6162307409 | -0.5974170774 | -1.557847309 |
| | 0.150 | -0.6848372934 | -0.8500952692 | -2.695002862 |
| 3s | 0.025 | -0.5298392842 | -0.3168335532 | -0.3091243145 |
| | 0.050 | -0.5696640174 | -0.4113242511 | -0.6050340761 |
| | 0.075 | -0.6173762645 | -0.5240248324 | -0.9557836844 |
| | 0.100 | -0.6713937028 | -0.6488631591 | -1.331376322 |
| | 0.150 | -0.7939098630 | -0.9203977116 | -2.099855330 |
| 3p | 0.025 | -0.5272195493 | -0.3128354399 | -0.3055088368 |
| | 0.050 | -0.5596966036 | -0.3986642162 | -0.6077677944 |
| | 0.075 | -0.5964276548 | -0.5009934106 | -0.9806451254 |
| | 0.100 | -0.6365166896 | -0.6146925374 | -1.388641432 |
| | 0.150 | -0.7238605000 | -0.8612543564 | -2.224098580 |
| 3d | 0.025 | -0.5244733721 | -0.3084948329 | -0.3011225088 |
| | 0.050 | -0.5487880544 | -0.3838064758 | -0.606747212 |
| | 0.075 | -0.5726108402 | -0.4717847684 | -0.9945628778 |
| | 0.100 | -0.5954927171 | -0.5678320312 | -1.419102816 |
| | 0.150 | -0.6365278213 | -0.7672835076 | -2.251887755 |
| 4s | 0.025 | -0.5323000579 | -0.3204668349 | -0.3120277371 |
| | 0.050 | -0.5785157811 | -0.4218033616 | -0.5996463159 |
| | 0.075 | -0.6350697965 | -0.5413329210 | -0.9254852509 |
| | 0.100 | -0.6995504645 | -0.6720273560 | -1.263244881 |
| | 0.150 | -0.8462213682 | -0.9524357106 | -1.935504414 |
| 4p | 0.025 | -0.5292122102 | -0.3161831200 | -0.3081839761 |
| | 0.050 | -0.5671318525 | -0.4084922092 | -0.5988699390 |
| | 0.075 | -0.6116097382 | -0.5170512012 | -0.9364831796 |
| | 0.100 | -0.6610134522 | -0.6354309818 | -1.288956165 |
| | 0.150 | -0.7700398596 | -0.8863990934 | -1.980709322 |

Table 2. Parameters of selected diatomic molecules used in this study [89]

| Molecules | $\mu(MeV)$ | $\alpha(\text{Å}^{-1})$ | $\mu(amu)$ |
|---|---|---|---|
| VH | 0.09203207571 | 1.44370 | 0.988005 |
| TiH | 0.09197301899 | 1.32408 | 0.987371 |
| TiC | 0.8948005221 | 1.52550 | 9.606079 |
| CuLi | 0.58306812793 | 1.00818 | 6.259494 |

Table 3. Bound state energy spectra $E_{nl}(eV)$ of the Eckart plus Hellmann potential for VH, TiH, TiC and CuLi diatomic molecules

| n | l | $E_{nl}$ (eV) of VH | $E_{nl}$ (eV) of TiH | $E_{nl}$ (eV) of TiC | $E_{nl}$ (eV) of CuLi |
|---|---|---|---|---|---|
| 0 | 0 | -4.388393324 | -3.995461324 | -4.687262254 | -3.014649864 |
| 0 | 1 | -4.363576420 | -3.976563694 | -4.684013156 | -3.013313002 |
| 0 | 2 | -4.315482766 | -3.939708307 | -4.677541581 | -3.010644546 |
| 0 | 3 | -4.246835875 | -3.886588443 | -4.667899981 | -3.006654929 |
| 0 | 4 | -4.160974946 | -3.819343691 | -4.655165101 | -3.001359541 |
| 0 | 5 | -4.061278595 | -3.740228732 | -4.639435781 | -2.994778429 |
| 1 | 0 | -4.351517256 | -3.974290474 | -4.665595573 | -3.015347165 |
| 1 | 1 | -4.329538507 | -3.957243671 | -4.662501668 | -3.014042854 |
| 1 | 2 | -4.286754140 | -3.923883123 | -4.656337945 | -3.011439114 |
| 1 | 3 | -4.225253356 | -3.875536963 | -4.647151876 | -3.007545617 |
| 1 | 4 | -4.147627936 | -3.813899190 | -4.635012960 | -3.002376632 |
| 1 | 5 | -4.056543954 | -3.740778524 | -4.620010768 | -2.995950756 |
| 2 | 0 | -4.330562184 | -3964109406 | -4.646784709 | -3.016648024 |
| 2 | 1 | -4.310755837 | -3.948515786 | -4.643830791 | -3.015373953 |
| 2 | 2 | -4.272055607 | -3.917910429 | -4.637944814 | -3.012830346 |
| 2 | 3 | -4.216096257 | -3.873352684 | -4.629169881 | -3.009026184 |
| 2 | 4 | -4.144926129 | -3.816204629 | -4.617569118 | -3.003974719 |
| 2 | 5 | -4.060681968 | -3.747937031 | -4.603223944 | -2.997693235 |
| 3 | 0 | -4.32179419 | -3.962583306 | -4.630535913 | -3.018511506 |
| 3 | 1 | -4.303698065 | -3.948159012 | -4.627708751 | -3.017265610 |
| 3 | 2 | -4.268227540 | -3.919778876 | -4.622074329 | -3.014778033 |
| 3 | 3 | -4.216682122 | -3.878299314 | -4.613671886 | -3.011057130 |
| 3 | 4 | -4.150704094 | -3.824829161 | -4.602558928 | -3.006115232 |
| 3 | 5 | -4.072027217 | -3.760577424 | -4.588809670 | -2.999968427 |
| 4 | 0 | -4.322668435 | -3.968078354 | -4.616595736 | -3.020901395 |
| 4 | 1 | -4.305952612 | -3.954615747 | -4.613883777 | -3.019681816 |
| 4 | 2 | -4.273099190 | -3.928071892 | -4.608478018 | -3.017246588 |
| 4 | 3 | -4.225154073 | -3.889146536 | -4.600414287 | -3.013603491 |
| 4 | 4 | -4.163448775 | -3.838751163 | -4.589745102 | -3.008764019 |
| 4 | 5 | -4.089403533 | -3.777888034 | -4.576538290 | -3.002743168 |
| 5 | 0 | -4.331402395 | -3.979423541 | -4.604744582 | -3.023785550 |
| 5 | 1 | -4.315825387 | -3.966767477 | -4.602137686 | -3.022590621 |
| 5 | 2 | -4.285139270 | -3.9051768464 | -4.596940511 | -3.020204428 |
| 5 | 3 | -4.240193658 | -3.905002862 | -4.589185844 | -3.016634230 |
| 5 | 4 | -4.182077111 | -3.857225419 | -4.578921771 | -3.011890752 |
| 5 | 5 | -4.111961137 | -3.799272206 | -4.566210430 | -3.005987997 |

Table 4. Comparison of energy eigenvalues $(eV)$ for a special case of Hellmann potential as a function of the screening parameter $\alpha$ with $\hbar = 2\mu = 1$ for $A = B = 0$, $B = 2$, and $D = -1$

| State | $\alpha$ | Present method | (NU) [76] | (AP) [77] | (PT) [78] |
|---|---|---|---|---|---|
| 1S | 0.001 | −2.250500250 | -2.250 500 | - 2.248 981 | - 2.249 00 |
| | 0.005 | −2.252506250 | -2.252 506 | - 2.244 993 | - 2.245 01 |
| | 0.01 | −2.255025000 | -2.255 025 | - 2.240 030 | - 2.240 05 |
| 2S | 0.001 | −0.5630010000 | - 0.563 001 | - 0.561 502 | - 0.561 502 |
| | 0.005 | −0.5650250000 | - 0.565 025 | - 0.557 549 | - 0.557 550 |
| | 0.01 | −0.5676000000 | - 0.567 600 | - 0.552 697 | - 0.552 697 |
| 2P | 0.001 | −0.5622502500 | - 0.563 000 | - 0.561 502 | - 0.561 502 |
| | 0.005 | −0.5612562500 | - 0.565 000 | - 0.557 541 | - 0.557 541 |
| | 0.01 | −0.5600250000 | - 0.567 500 | - 0.552 664 | -0.552 664 |
| 3S | 0.001 | −0.2505022500 | - 0.250 502 | - 0.249 004 | - 0.249 004 |
| | 0.005 | −0.2525562500 | - 0.252 556 | -0245 110 | - 0.245 111 |
| | 0.01 | −0.2552250000 | - 0.255 225 | -0.240 435 | - 0.240 435 |
| 3p | 0.001 | −0.2501680278 | -0.250 501 | - 0.249 004 | - 0.249 004 |
| | 0.005 | −0.2508673611 | -0.252 531 | -0.245 102 | -0.245 103 |
| | 0.01 | −0.2518027778 | -0.255 125 | -0.240 404 | -0:240 404 |
| 3d | 0.001 | −0.2495002500 | -0.250 833 | -0.249 003 | -0.249 003 |
| | 0.005 | −0.2475062500 | -0.254 151 | -0.245 086 | -0.245 086 |
| | 0.01 | −0.2450250000 | -0.258 269 | -0.240 341 | -0.240 341 |
| 4S | 0.001 | −0.1411290000 | -0.141 129 | -0.139 633 | -0.139 633 |

| | 0.005 | −0.1432250000 | -0.143 225 | -0.135 819 | -0.135 819 |
| | 0.01 | −0.1460250000 | -0.146 025 | -0.131 380 | -0.131 381 |
| 4p | 0.001 | −0.1409405625 | -0.141 128 | -0.139 632 | 0.139 633 |
| | 0.005 | −0.1422640625 | -0.143 200 | -0.135 811 | 0.135 811 |
| | 0.01 | −0.1440562500 | -0.145 925 | -0.131 350 | -0.131 351 |
| 4d | 0.001 | −0.1405640625 | -0.141 314 | -0.139 632 | -0.139 632 |
| | 0.005 | −0.1403515625 | -0.144 089 | -0.135 795 | -0.135 796 |
| | 0.01 | −0.1401562500 | -0.147 606 | -0.131 290 | -0.131 290 |
| 4f | 0.001 | −0.1400002500 | -0.141 686 | -0.139 631 | -0.139 631 |
| | 0.005 | −0.1375062500 | -0.145 902 | -0.135 772 | -0.135 772 |
| | 0.01 | −0.1344000000 | -0.151 106 | -0.131 200 | -0.131 200 |

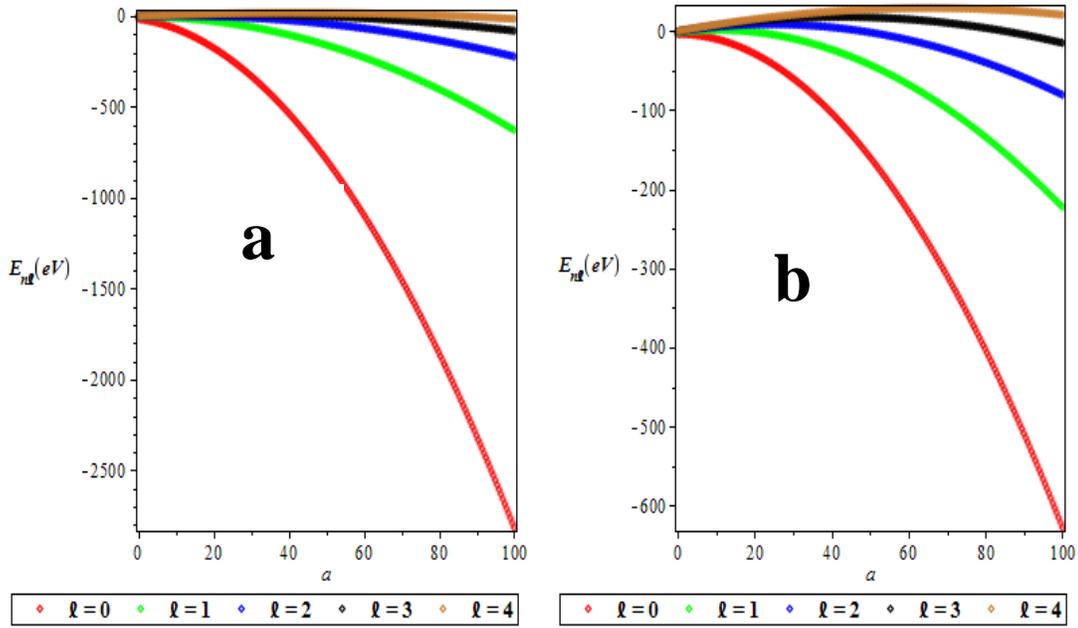

FIGURE 1(a). Variation of the ground state energy spectra for various $l$ as a function of $a$. (b). The plot of the first excited state energy spectra for different $l$ as a function of $a$. We choose $A = 1$, $B = -1$, $C = 4$, $D = -4$ and $\alpha = 0.025$ for the ground and excited states.

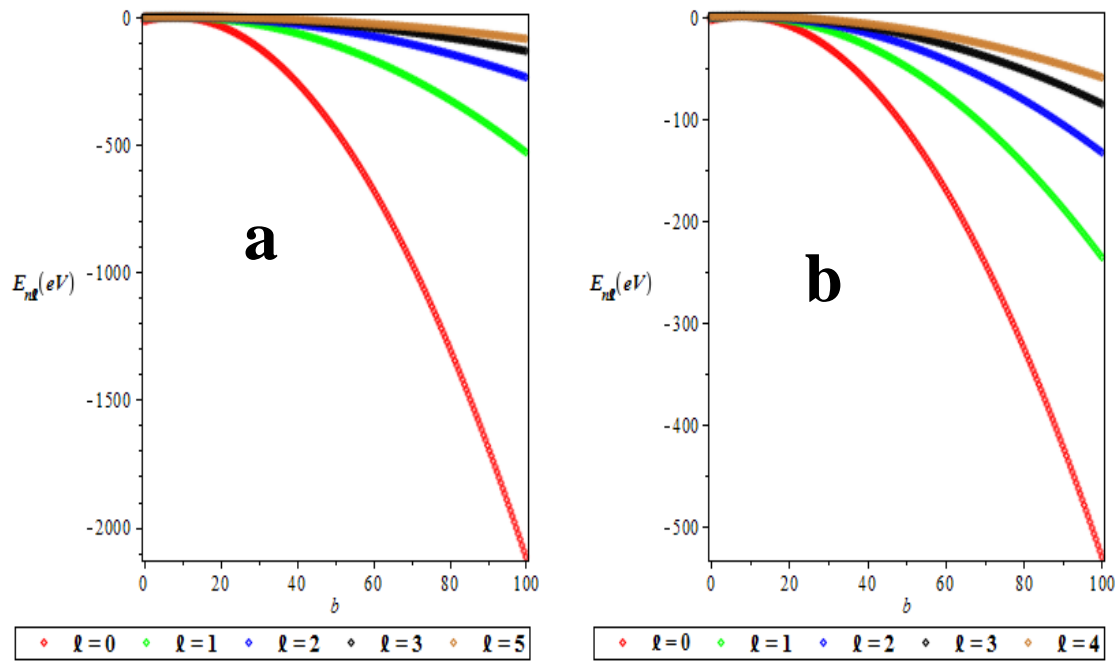

FIGURE 2(a). Variation of the ground state energy spectra for various $l$ as a function of $b$. (b) A plot of the first excited state energy spectra for various $l$ as a function of $b$. We choose $A=1$, $B=-1$, $C=4$, $D=-4$ and $\alpha=0.025$ for the ground and excited states.

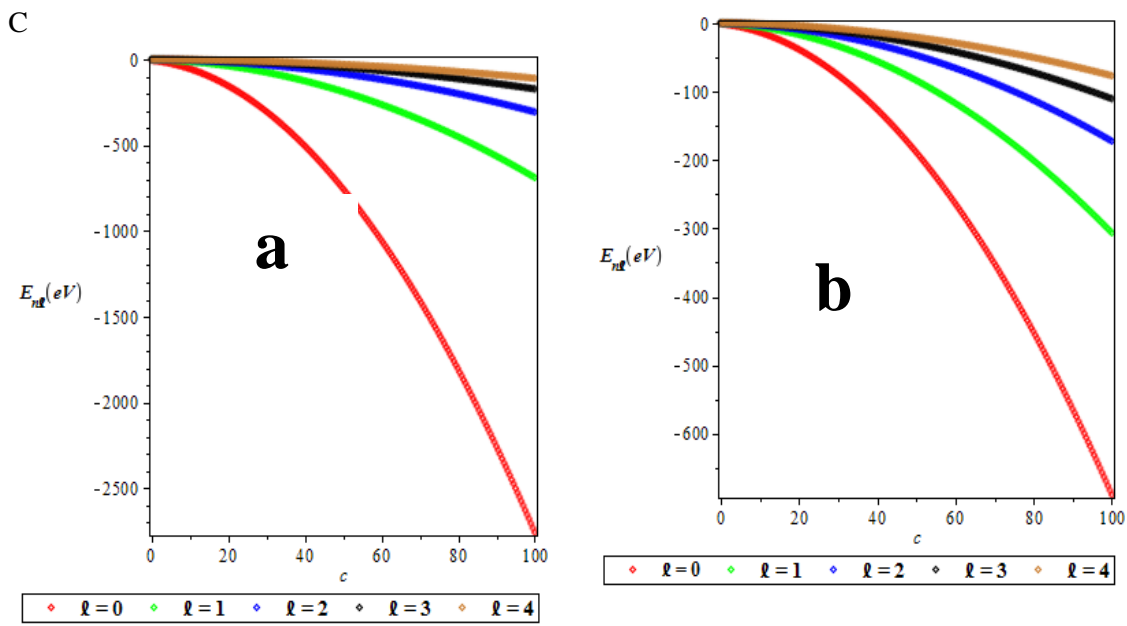

FIGURE 3(a). Variation of the ground state energy spectra for various $l$ as a function of $c$.
(b). The plot of the first excited state energy spectra for various $l$ as a function of $c$. We choose $A=1$, $B=-1$, $C=4$, $D=-4$ and $\alpha=0.025$ for the ground and excited states

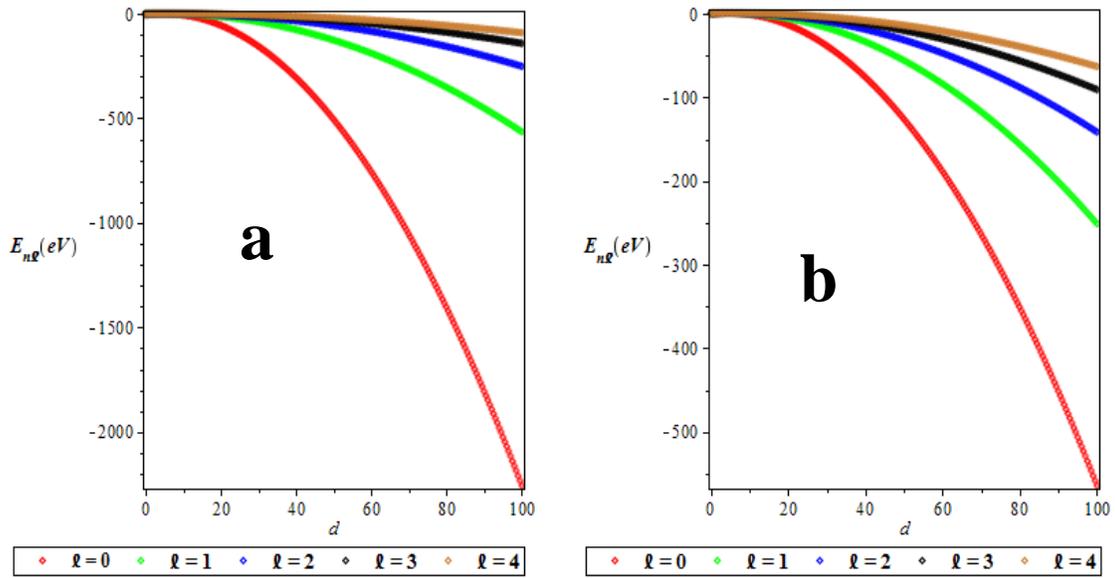

FIGURE 4(a). Variation of the ground state energy spectra for various $l$ as a function of $d$.
(b). The plot of the first excited state energy spectra for various $l$ as a function of $d$. We choose $A = 1,\ B = -1,\ C = 4,\ D = -4$ and $\alpha = 0.025$ for the ground and excited states

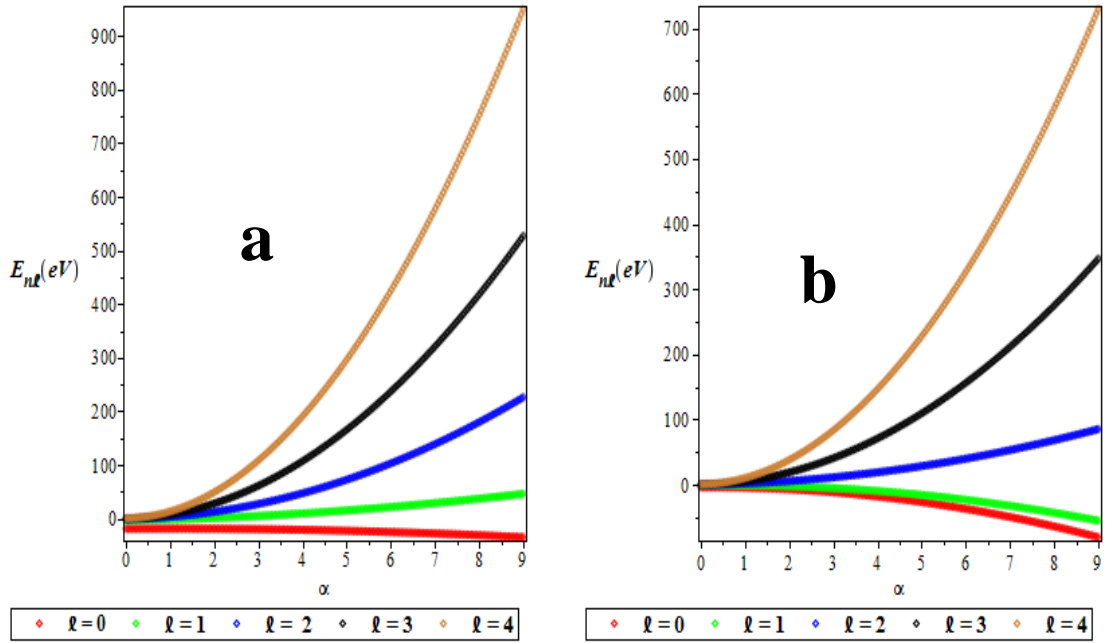

FIGURE 5(a). Variation of the ground state energy spectra for various $l$ as a function of the screening parameter $(\alpha)$ (b). A plot of the first excited state energy spectra for different $l$ as a function of the screening parameter $(\alpha)$. We choose $A = 1,\ B = -1,\ C = 4,\ D = -4$ and $\alpha = 0.025$ for the ground and excited states

## 5. Conclusion

In this research, the bound state solutions to the Schrödinger equation with EHP have been studied within the Greene-Aldrich approximation scheme. The eigenvalues and the eigen functions are obtained using the NUFA method. We then apply the energy equation for four diatomic molecules by imputing the experimental values of each molecular parameter. The results show that the bound state energy spectra of these diatomic molecules increases as various quantum numbers $n$ and $l$ increase. To test the accuracy of our results, we computed the bound states energy (eV) eigenvalues of EHP and Hellmann potential which agree with the report of other researchers. We plotted the ground and excited states energy eigenvalues of the different quantum states as a function of the EHP strengths, respectively. We observed that there is a decrease in energy in both the ground and excited states as the potential strength increases.

.